\documentclass[aps,prl,preprint,showpacs,superscriptaddress,amsmath,amssym,nofootinbib]{revtex4}
\usepackage{graphicx}
\usepackage{dcolumn}
\usepackage{bm}
\usepackage{color}

\begin{document}


\title{$^{11}$B and Constraints on Neutrino Oscillations and Spectra from Neutrino Nucleosynthesis}
\author{Sam M. Austin}
\email{austin@nscl.msu.edu}
\homepage{www.nscl.msu.edu/~austin}
\affiliation{National Superconducting Cyclotron Laboratory,\\
1 Cyclotron, Michigan State University, East Lansing, MI 48824-1321\\
Joint Institute for Nuclear Astrophysics}
\affiliation{Department of Physics and Astronomy, Michigan State
University, East Lansing Michigan 48824}
\author{Alexander~Heger}
\affiliation{School of Physics and Astronomy,
University of Minnesota, Twin Cities,
Minneapolis, MN 55455-0149\\Joint Institute for Nuclear Astrophysics}
\email{alex@physics.umn.edu}
\author{Clarisse~Tur}
\affiliation{National Superconducting Cyclotron Laboratory,\\
1 Cyclotron, Michigan State University, East Lansing, MI 48824-1321\\
Joint Institute for Nuclear Astrophysics}

\date{\today}
\begin{abstract}
  We have studied the sensitivity to variations in the triple alpha
  and $^{12}$C($\alpha, \gamma$)$^{16}$O reaction rates, of the yield
  of the neutrino process isotopes $^7$Li, $^{11}$B, $^{19}$F,
  $^{138}$La, and $^{180}$Ta in core collapse supernovae. Compared
  to solar abundances, less than 15\% of
  $^7$Li, about 25-80\% of $^{19}$F, and about half of $^{138}$La
  is produced in these stars.  Over a range of $\pm
  2\sigma$ for each helium-burning rate, $^{11}$B is
  overproduced and the yield varies by an amount larger than the
  variation caused by the effects of neutrino oscillations. The
  total $^{11}$B yield, however, may eventually provide
  constraints on supernova neutrino spectra.
\end{abstract}


\maketitle


About $10^{58}$ neutrinos are emitted during a typical core collapse
supernova explosion.  For some time it has been known (see
\cite{woo90} for a detailed history) that interactions of these
neutrinos with the stellar envelope can produce certain rare nuclei in
abundances close to those observed in nature.  These nuclei, called
here the neutrino nuclei, include $^7$Li, $^{11}$B, $^{19}$F,
$^{138}$La, and $^{180}$Ta \cite{woo90,heg05}.

It was pointed out \cite{heg05} that the production of some of the
$^{180}$Ta and most of the $^{138}$La by the neutrino
process was sensitive to the electron neutrino temperatures, and
might serve to probe the value of the neutrino oscillation parameter
$\sin^22\theta_{13}$. Recently, \cite{yos06a,yos06b,kaj08} showed that
the yields of $^7$Li and $^{11}$B in supernova explosions are also
sensitive to $\sin^22\theta_{13}$ and to whether the neutrino mass
hierarchy is normal or inverted.  In both cases, this sensitivity
arises because neutrino oscillations can change the neutrino spectra
produced during core collapse supernovae, increasing the average
energies of the $\nu_e$ and $\bar{\nu}_e$ and affecting the synthesis
of the neutrino nuclei.  Since two of the main goals of neutrino
physics \cite{p508} are to determine better the value of
$\sin^22\theta_{13}$ and the nature of the mass hierarchy, the
possibility that the observed abundances of the neutrino nuclei might
constrain these quantities is of great interest.

Their use for this purpose depends, however, on the robustness of the stellar yield predictions.
Studies of the dependence of nucleosynthesis on the helium burning
reaction rates have shown \cite{tur07, tur09} that both the yields of the more abundant nuclides
and stellar structure are significantly affected.  Since the neutrino
nuclei result from neutrino induced spallation of abundant progenitor nuclei,
their production depends on the abundances of these nuclei and on
their location within the star, and thereby on the rates of the helium burning reactions.

In  this paper, we examine the changes in the production of $^7$Li, $^{11}$B,
$^{19}$F, $^{138}$La, and $^{180}$Ta caused by changes in the
astrophysical helium burning rates within their uncertainty limits,
and compare the yield changes of $^7$Li, and $^{11}$B, with the
predicted \cite{yos06a,yos06b,kaj08} effects of oscillations.

We then
discuss how, and whether, the neutrino process nuclei can be used to constrain the
neutrino spectra from supernovae. We find that the constraints provided by neutrino process
nucleosynthesis are interesting but not yet definitive.  Because of
the great interest in these issues it appears that a major effort to
sharpen these constraints is warranted; a discussion of important
measurements and calculations is given below.

We used the KEPLER code \cite{wea78,woo95,rau02,woo02} to model the
evolution of 15, 20, and 25 solar mass stars from central hydrogen
burning up to core-collapse; a piston placed at the base of the oxygen
shell was then used to simulate the explosion.  Following \cite{heg05}
we assumed a total energy of $5\times10^{52}$ ergs per neutrino
species, i.e. a total of $3\times10^{53}$ ergs energy release in the supernova
explosion.  Mass loss processes were included.
The neutrino spectra
were approximated by Fermi-Dirac distributions with a zero degeneracy
parameter, a luminosity exponentially decaying after onset of core
collapse with a time-scale of 3 s and a constant neutrino temperature:
$T=4$ MeV for $\nu_e$ and $\bar{\nu}_e$; $T=6$ MeV~for
${\nu}_{\mu},~\bar{\nu}_{\mu},~\nu_{\tau}, and~\bar{\nu}_{\tau}$. For
further details, see \cite{tur07, tur09, tur10, heg05}.  These choices are consistent with estimates of neutrino emission intensity and time dependence from supernovae \cite{arn89}.

Initial stellar abundances were taken from both Anders \& Grevesse
\cite{and89} and from Lodders \cite{lod03}, hereafter AG89 and L03. The
L03 abundances for C, N, O, Ne are roughly 15\%-25\% lower than those
of AG89, whereas the abundances of heavier elements are roughly 15\%
higher.
For calculations of neutrino process
  cross sections we used the results from \cite{heg05}.  Briefly, in
  that paper the charged and neutral current cross sections were first
  used to calculate the excitation spectra of the product nuclei;
  experimental data and $0 \hbar\omega$ shell-model estimates were
  used to determine the Gamow-Teller response for $^{12}$C and
  $^{20}$Ne (leading to $^{11}$B and $^{19}$F) and RPA estimates for
  the $J\le4$ multipoles for all other transitions.  The SMOKER
  statistical model code \cite{rau02}, was then used there to follow
  the ensuing decays.  The $\gamma $ process contributions to the
yields are also included in our calculations, but are important only
for $^{180}$Ta.

\begin{figure}
\centering
\includegraphics[width=8. cm]{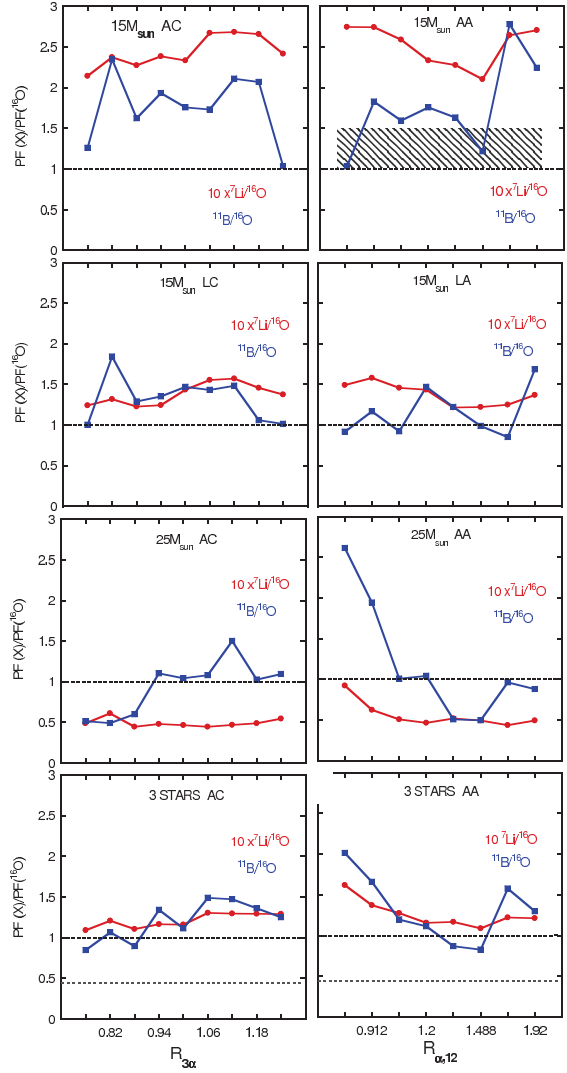}
\caption{Production factors of $^7$Li and $^{11}$B compared to those of
  $^{16}$O for various reaction rates.  The left hand column shows the
  results when the triple-alpha reaction rate $R_{3\alpha}$ is varied
  about the central value of $1.0$, the rate of ref. \cite{cau88}.  The
  right-hand column shows the results when $R_{\alpha,12}$ is varied about
  the central value of $1.2$.  The value for $^7$Li has been multiplied
  by a factor of $10$. An example of the range of variation in $^{11}$B yield
  predicted in \cite{yos06a}
is shown as a band in the upper right-hand panel.
  The dotted line at 0.4 is the production factor ratio that would
  give the solar abundance of $^{11}$B not made in the galactic cosmic rays. For more information see the
  text.}
\end{figure}

Isotope yields were stored at nine key points
of stellar evolution \cite{tur09,tur10}.  As anticipated, $^{7}$Li and $^{11}$B
were produced essentially only during the supernova stage;  their yields are shown in Fig.~1.
Here an initial ``A'' (or ``L'') label means that the calculations
were done for the AG89 (or L03) abundances, and a final
``A'' (or ``C'') means that the $^{12}$C($\alpha, \gamma$)$^{16}$O
(or triple-alpha) rates were varied by $\pm 2\sigma$ from their
central values.  These central values were, resp., $1.2$ times the
rate recommended by Buchmann \cite{buc96} with $\sigma = 25\%$ and
that recommended by Caughlan and Fowler \cite{cau88} with $\sigma =
12\%$. For the $^{12}$C($\alpha, \gamma$)$^{16}$O
rate, the central value is that commonly used in calculations with the KEPLER
code \cite{woo07,tur07}; it is consistent with recent measurements \cite{tan10}. The energy dependence obtained by Buchmann was used for all calculations.
The labels also give the stellar mass, or 3 STARS, the average
for the 15, 20, and 25 $\mathrm{M}_\odot$ stars using a Scalo \cite{sca86}
Initial Mass Function (IMF) with a slope of $\gamma = -2.65$. For
normalization purposes we  compare to the production factor
for oxygen. Since
$^{16}$O is made mainly in massive stars, a production
factor ratio near one is consistent with all (or most) of an isotope
being made in a primary neutrino process (as are $^7$Li, $^{11}$B,
and $^{19}$F).

Examining first the results for average production in the three-star
sample, and assuming that this is a reasonable approximation of the
total production process, we see that only 10\%-15\% of $^7$Li is made
in the neutrino process.  This is not surprising, since there
are many processes, including the Big Bang, that make or destroy
$^7$Li and that are not fully understood. On the other hand, for most
values of the reaction rates $^{11}$B is overproduced, even if one ignores production by cosmic rays.

Fig.~2 shows the results for $^{138}$La and $^{180}$Ta.  Here we show
also results for the pre-SN stage, the time when the contraction speed in the
iron core reaches 1000 km sec$^{-1}$, since production
during that stage is not negligible, especially for $^{180}$Ta. A
detailed examination, however, shows that most of the $^{138}$La
and $^{180}$Ta that is ejected in the SN is not what was
present in the pre-SN stage; that is mostly
destroyed by the SN shock and most of what is ejected was newly
synthesized during the explosion \cite{rau02,heg05}.

Since these two isotopes are secondary products (produced from pre-existing spallation targets)
a production ratio of about two for $^{138}$La and
$^{180}$Ta (see the dotted line on Fig.~2) would be necessary to reproduce the solar abundance. An additional complication is that our
models do not distinguish production in the short lived
ground state from that in the long-lived isomeric $9^-$ state in $^{180}$Ta;
a better, but still approximate, treatment \cite{hay10} gives an isomer production of about 40\% of the total production. It then appears that the production of $^{180}$Ta is roughly consistent with the solar abundance, given the uncertainties in the production calculations, and that the production of $^{138}$La corresponds to about half the solar abundance. $^{19}$F is
a primary product, and it appears that 25\%-75\% of solar $^{19}$F could be made by the
neutrino process.  This   complicates the determination of the
importance of other sources such as AGB stars and Wolf-Rayet winds.

\begin{figure}
\centering
\includegraphics[width=8 cm]{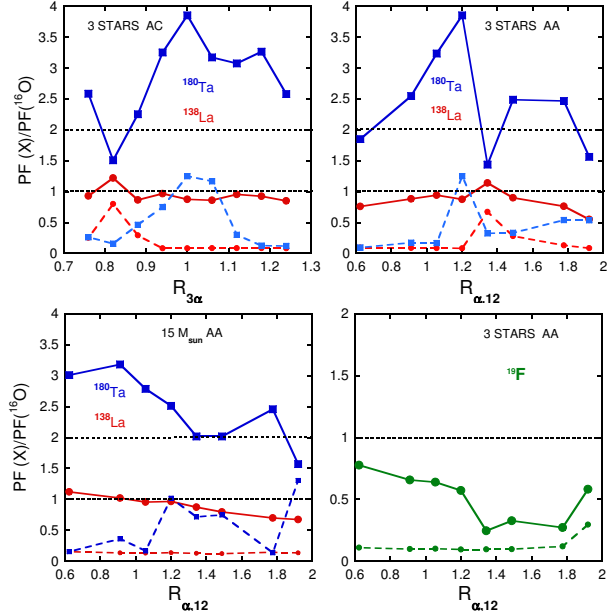}
\caption{(Top left): Production factors (three star averages) for
  $^{138}$La and $^{180}$Ta using Anders and Grevesse abundances and
  varying the triple alpha rate.  (Top right): Same but varying
  $R_{\alpha,12}$. (Lower left): Same for 15 $\mathrm{M}_\odot$ star, varying
  $R_{\alpha,12}$.  (Lower right): Production factors (3 star average)
  for$^{19}$F. The solid curves are for the final abundances and the
  dashed curves are for the pre-SN stage.  All production factors are
  ratios to that of $^{16}$O.}
\end{figure}

We now consider whether a comparison of the observed
abundances of $^7$Li and $^{11}$B to SN model predictions can place
constraints on the neutrino oscillation process, as was suggested in
Refs.~\cite{yos06a, yos06b,kaj08}.  These investigations were for a
16.2 $\mathrm{M}_\odot$ star, using parameters almost identical to those we
have used, except that the explosion energy, $T_{\nu_e}$, and $T_{\bar
  \nu_e}$ were 1.0 Bethe, 3.2 MeV, and 5.0 MeV instead of 1.2 Bethe, 4.0
MeV and 4.0 MeV.  The near equality of the average
neutrino energies should yield similar production for the two models
in the absence of neutrino oscillations.

In \cite{yos06a, yos06b,kaj08} neutrino oscillations produce
significant increases in $^7$Li production, up to 75\%, as
$\sin^22\theta_{13}$ increases from $10^{-6}$ to $10^{-2}$ for the
normal neutrino hierarchy--the changes are much smaller, around 15\%
for an inverted hierarchy. The changes for $^{11}$B are small 
for either hierarchy, around 20\%. The number ratio $N(^7$Li
)/$N(^{11}$B) is assumed to be less susceptible then the absolute
yields, to systematic uncertainties in the calculations and has
approximately 50\% changes for the normal hierarchy, and less than 10\% for the
inverted hierarchy. This provides, in principle, some hope that
observation of an enhanced ratio could place a lower limit on the
value of $\sin^22\theta_{13}$ and eliminate the option of an inverted
hierarchy.

The variations with the helium burning reaction rates, of the production
of $^7$Li by the neutrino process are relatively small, about
20\%, but it will be difficult to untangle the relatively small amount
of neutrino-produced $^7$Li from other sources of $^7$Li.  One might
hope that observations of $^7$Li in pre-solar grains would make it
possible to isolate the effects of individual supernovae. Unfortunately, few, if any,
relevant observations have been made to date.  Lithium
isotopic ratios have been measured in very large SiC grains by
Gyngard, et al. \cite{gyn09}, but Li is very volatile and is not
expected to condense into SN SiC grains \cite{nit10}. Thus, while its production does not depend strongly on the helium burning rates,
other considerations limit its usefulness.

The situation for $^{11}$B is also unclear.  Hoppe et al.~\cite{hop01}
measured B in supernova SiC grains.  Their measured isotopic ratio,
$^{11}$B/$^{10}$B $= 3.46 \pm 1.36$, is, however, consistent with
laboratory contamination by solar system B ($^{11}$B/$^{10}$B$ = 4.045$).
They conclude that at most 30\% of the measured B is attributable to the
neutrino process. Their Fig.~5 \cite{hop01} shows a corrected value
reduced by a corresponding factor of three. After this correction,
the abundance of B is lower than expected, by over an
order of magnitude.  They consider some possible explanations, but
it appears that the grain formation process is not well understood.

Moreover, as we see in the top four panels of Fig.~1, variations in $^{11}$B yields with
reaction rate are large, a factor of two or more, making the
uncertainties in its predicted ratio to $^7$Li and in its absolute
value much larger than the effects predicted by \cite{yos06a,yos06b,kaj08}.
We conclude that this approach to constraining neutrino oscillations
will not be productive until there are significant improvements in the
helium burning rates discussed here, as well as in grain observations and
their interpretation.

Before considering whether the gross production rates
might eventually provide a constraint on neutrino spectra,
we need to examine the total production of $^{11}$B. Both $^{11}$B and $^{10}$B are
made in the galactic cosmic rays
(GCR) with the ratio $^{11}$B/$^{10}$B lying between 2.2 and 2.5 \cite{men71,pra10a}. The $\pm 0.15$
uncertainties in the ratio reflect, mainly, uncertainties in the cosmic ray sources
and the propagation model. (For reviews see
\cite{pra07,pra10}).  We take as the observed meteoritic ratio 4.045 \cite{lod03};
the other recent abundance summaries \cite{and89,lod10} quote results
within 0.5\% of this value. Since we find that the neutrino process makes little $^{10}$B
 ($^{11}$B/$^{10}$B $\approx 50$), about $42 \pm4$\% of the $^{11}$B must
 be made in the neutrino process. A possible contribution of (so far  unobserved) low
energy cosmic rays would increase the $^{11}$B/$^{10}$B ratio in
cosmic ray production \cite{men71} but it has been found \cite{ram97} that this process is
energy inefficient and unlikely to produce a significant amount of $^{11}$B.  The neutrino process $^{11}$B
should then be compared to about $0.4 \times$(solar $^{11}$B)--this comparison is made
in Fig.~1.

Summarizing, except for $^7$Li and $^{19}$F, which have other known production sites,
it appears that production by the neutrino process (and partially, for  $^{180}$Ta
by the gamma process), as shown in the 3-Star panels of Figs. 1 and 2,
is within a factor of three of the observed abundances except
for extreme values of the rates.

It has
been shown previously that neutrino process yields increase strongly
for larger neutrino energies. For example, increasing the temperature
of $\nu_e$ and $\bar{\nu}_e$ neutrinos from 4 to 6 MeV increases the
yields of the neutrino nuclei by factors from 1.5 to 2
\cite{heg05}. Similar changes were obtained for ${\nu}_{\mu}$.  This
strong dependence raises the possibility of constraining the ranges of allowable neutrino
temperatures, spectral shapes and neutrino intensity. Such constraints depend on the robustness of the model predictions, and
thereby on the nuclear rates, on the neutrino interaction cross
sections, on the form of the neutrino spectra, and on the
astrophysical modeling uncertainties of the underlying stellar
models. It is probable that the best limits will be obtained for
$^{11}$B. The neutrino interaction cross sections for $^{12}$C can
be more reliably calculated than those for the heavier nuclei,
because the strong Gamow-Teller cross sections are mainly experimentally based,
and shell model estimates can  replace RPA calculations
for the $L>0$ cross sections \cite{aue02,vol00,hay00}. The constraint imposed by the
meteoritic and GCR $^{11}$B/$^{10}$B ratios is also useful in
determining the appropriate SN $^{11}$B yield.

Taken at face value, it seems that significantly
harder neutrino spectra than we have used are improbable--the yield is already overestimated.
But improvements in the neutrino process calculations are necessary to make this constraint credible.
It appears likely \cite{aus10} that the uncertainty in the triple alpha
rate will be halved in the near future and there are major efforts to improve the $^{12}$C($\alpha,\gamma$)$^{16}$O rate.  Better
estimates of neutrino spectra can also be employed; it is now known
\cite{jan07,lan08} that the mean energies of the various neutrino
species are more similar than had been thought, and that the high
energy tail is suppressed by inclusion of inelastic scattering
processes. The mean energies and second energy moments of these new spectra
are, however, similar to those of the Fermi-Dirac distributions we have used,
differing by less than 12 \% in all cases-the second moments are related to the neutrino process cross sections.
The astrophysical model uncertainties are a remaining
difficulty, but these should also be reduced using techniques informed
by 3-D calculations.


To summarize, we explored changes in the core-collapse supernova
yields of $^7$Li, $^{11}$B, $^{19}$F, $^{138}$La, and $^{180}$Ta that
arise from changes in the triple alpha and
$^{12}$C($\alpha,\gamma$)$^{16}$O reaction rates within their $\pm 2\sigma$ uncertainties.  We found that
the rate changes result in factor of two changes in the
production of $^{11}$B in a 15 $\mathrm{M}_\odot$ star. This, for the present at least, rules out the
techniques proposed \cite{yos06a,yos06b,kaj08} to constrain the
neutrino oscillation parameter $\sin^22\theta_{13}$. For the assumed neutrino spectra there is
significant overproduction of $^{11}$B for all values of the
rates we have used ($\pm 2\sigma$).  It seems reasonable to expect that a
factor of two improvement in the precision of neutrino-process nucleosynthesis
can be achieved, especially for $^{11}$B. This may provide a constraint on the neutrino energy spectrum.
If one assumes that model calculations can
accurately fix the spectral shape, neutrino process
nucleosynthesis could provide an estimate of the neutrino flux from
supernovae and a check on supernova models that does not depend on
occurrence of (infrequent) supernova explosions.


We thank Robert Hoffman and Stan Woosley for assistance with reaction
rates and for helpful discussions. Research support from: US NSF: grants PHY06-06007, PHY02-16783(JINA)); US
DOE: contract DE-AC52-06NA25396, grants DE-FC02-01ER41176, FC02-09ER41618 (SciDAC), DE-FG02-87ER40328.

\end{document}